%% file: fracture-harmonics.tex
\PassOptionsToPackage{table}{xcolor}
\documentclass[acmtog,nonacm,balance=false]{acmart}
\setcitestyle{square}

\input{common-preamble.tex}

\renewcommand{\k}{k}
\hypersetup{draft}
\begin{document} %

\title{Fracture Modes for Realtime Destruction}

\author{Silvia Sell\'{a}n}
\affiliation{\institution{University of Toronto}}

\author{Jack Luong}
\affiliation{\institution{California State University,
Fresno and University of California, Los Angeles}}

\author{Leticia Mattos Da Silva}
\affiliation{\institution{University of California, Los Angeles and
Massachusetts Institute of Technology}}

\author{Aravind Ramakrishnan}
\affiliation{\institution{University of Maryland at College Park and University
of Toronto}}

\author{Yuchuan Yang}
\affiliation{\institution{University of California, Los Angeles}}

\author{Alec Jacobson}
\affiliation{\institution{University of Toronto and Adobe Research}}

\begin{abstract}
  \input{sections/abstract.tex}
\end{abstract}

\begin{teaserfigure}
 \centering
 \vspace{-0.4cm}
 \includegraphics{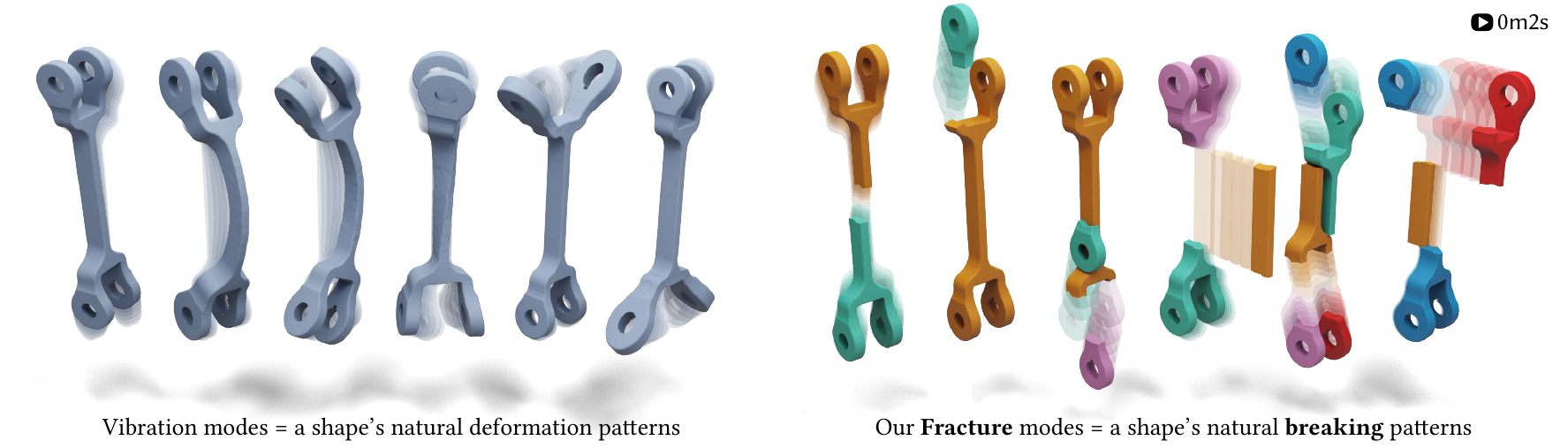} 
 \vspace{-0.7cm}
 \caption{Drawing an analogy with the well-studied vibration modes, we define a
 shape's \emph{fracture modes}, which we can precompute for realtime applications.
  } 
\label{fig:modes-vs-modes}
\end{teaserfigure}

\maketitle
\vspace{-0.1cm}
\input{sections/intro.tex}

\input{sections/related.tex}

\input{sections/method.tex}

\input{sections/results.tex}
\input{sections/conclusion.tex}
\input{sections/acks.tex}

\bibliographystyle{ACM-Reference-Format}
\bibliography{references.bib}
\input{sections/appendix}
\end{document}

%% file: common-preamble.tex
\usepackage[utf8]{inputenc}
\usepackage[verbose]{newunicodechar}
\newcommand{\wf}[6]{%
  \vspace*{-0.75\intextsep}
  \hspace*{-0.5\columnsep}
  \begin{wrapfigure}[#1]{#2}{#3}%
  \centering%
  \includegraphics[#4]{#5}
  #6%
\end{wrapfigure}}
\definecolor{derekTableBlue}{rgb}{0.565,0.847,0.769} 
\newunicodechar{Δ}{\Delta}
\newunicodechar{Γ}{\Gamma}
\newunicodechar{Λ}{\Lambda}
\newunicodechar{Ω}{\Omega}
\newunicodechar{Φ}{\Phi}
\newunicodechar{Π}{\Pi}
\newunicodechar{Ψ}{\Psi}
\newunicodechar{Σ}{\Sigma}
\newunicodechar{∑}{\sum}
\newunicodechar{σ}{\Sigma}
\newunicodechar{Θ}{\Theta}
\newunicodechar{Υ}{\Upsilon}
\newunicodechar{Χ}{\Xi}
\newunicodechar{α}{\alpha}
\newunicodechar{β}{\beta}
\newunicodechar{ξ}{\chi}
\newunicodechar{δ}{\delta}
\newunicodechar{ε}{\epsilon}
\newunicodechar{η}{\eta}
\newunicodechar{γ}{\gamma}
\newunicodechar{ι}{\iota}
\newunicodechar{κ}{\kappa}
\newunicodechar{λ}{\lambda}
\newunicodechar{μ}{\mu}
\newunicodechar{ν}{\nu}
\newunicodechar{ω}{\omega}
\newunicodechar{φ}{\varphi}
\newunicodechar{π}{\pi}
\newunicodechar{ψ}{\psi}
\newunicodechar{ρ}{\rho}
\newunicodechar{σ}{\sigma}
\newunicodechar{τ}{\tau}
\newunicodechar{θ}{\theta}
\newunicodechar{υ}{\upsilon}
\newunicodechar{χ}{\xi}
\newunicodechar{ζ}{\zeta}

\newcommand{\ones}{\mathbf{1}}
\newcommand{\zeros}{\mathbf{0}}

\newunicodechar{ℝ}{\mathbb{R}}
\newunicodechar{ᵀ}{^\top}
\newunicodechar{→}{\rightarrow}
\newunicodechar{↓}{\downarrow}
\newunicodechar{∈}{\in}
\newunicodechar{ⁿ}{^{n}}
\newunicodechar{₂}{_{2}}
\newunicodechar{×}{\times}
\newunicodechar{∀}{\forall}
\newunicodechar{∞}{\infty}

\newcommand{\argmin}{\mathop{\text{argmin}}}

\newcommand{\trace}[1]{\mathop{\text{trace}}\left(#1\right)}

\newcommand{\new}[1]{#1}
\newcommand{\minor}[1]{#1}

\newcommand{\refequ}[1]{Eq.~(\ref{equ:#1})}
\newcommand{\refapp}[1]{Appendix \ref{app:#1}}
\newcommand{\reffig}[1]{Fig.~\ref{fig:#1}}
\newcommand{\refsec}[1]{Section~\ref{sec:#1}}
\newcommand{\refalg}[1]{Algorithm~\ref{alg:#1}}

\providecommand{\C}{}
\providecommand{\D}{}

\providecommand{\I}{}

\providecommand{\L}{}
\providecommand{\M}{}

\providecommand{\Q}{}
\providecommand{\R}{}

\providecommand{\U}{}

\providecommand{\X}{}
\providecommand{\Y}{}
\providecommand{\Z}{}

\providecommand{\c}{}

\providecommand{\k}{}

\providecommand{\u}{}
\providecommand{\v}{}
\providecommand{\w}{}
\providecommand{\x}{}

\providecommand{\z}{}

\renewcommand{\D}{\mathbf{D}}
\renewcommand{\M}{\mathbf{M}}
\renewcommand{\I}{\mathbf{I}}
\renewcommand{\U}{\mathbf{U}}
\renewcommand{\Q}{\mathbf{Q}}

\renewcommand{\R}{\mathbb{R}}

\renewcommand{\Y}{\mathbf{Y}}
\renewcommand{\L}{\mathbf{L}}
\renewcommand{\u}{\mathbf{u}}
\renewcommand{\v}{\mathbf{v}}
\renewcommand{\w}{\mathbf{w}}
\renewcommand{\x}{\mathbf{x}}

\renewcommand{\z}{\mathbf{z}}
\renewcommand{\c}{\mathbf{c}}
\renewcommand{\C}{\mathbf{C}}
\renewcommand{\Z}{\mathbf{Z}}
\renewcommand{\X}{\mathbf{X}}

\usepackage{tabularx}
\usepackage{booktabs}
\usepackage{makecell}
\usepackage{float}
\usepackage[table]{xcolor}
\definecolor{white}{rgb}{1,1,1}
\definecolor{lightbluishgrey}{rgb}{0.76471,0.84824,0.91647}

\usepackage[ruled,vlined]{algorithm2e}
\SetKwRepeat{Do}{do}{while}
\SetKwComment{Comment}{\quad //}{}

\usepackage{subcaption}
\usepackage{wrapfig}
\usepackage{graphicx}
\usepackage{listings}

\usepackage{layouts}
\makeatletter 
\newcommand{\layoutdetails}{%
\begin{tabular}{ll}
 \texttt{\textbackslash{textwidth}} & \printinunitsof{in}\prntlen{\textwidth} \\
\texttt{\textbackslash{linewidth}} & \printinunitsof{in}\prntlen{\linewidth} \\
Main text font &  \f@size pt \f@family \\
\sffamily \small Caption text font &  \sffamily \small \f@size pt \f@family \\
\end{tabular}%
}
\makeatother 

\theoremstyle{remark}

\theoremstyle{remark}

%% file: sections/abstract.tex
Drawing a direct analogy with the well-studied vibration or elastic modes, we
introduce an object's \emph{fracture modes}, which constitute its preferred or
most natural ways of breaking. We formulate a sparsified eigenvalue problem,
which we solve iteratively to obtain the $n$ lowest-energy modes. These can be
precomputed for a given shape to obtain a prefracture pattern that can
substitute the state of the art for realtime applications at no runtime cost but
significantly greater realism.
Furthermore, any realtime impact can be projected onto our modes to obtain
impact-dependent fracture patterns without the need for any online crack propagation
simulation. We not only introduce this theoretically novel concept, but also
show its fundamental and practical \minor{advantages} in a diverse set of examples and
contexts.

%% file: sections/intro.tex
\section{Introduction}\label{sec:intro}

\begin{figure}[b!]
  \centering
  \vspace{-0.4cm}
  \includegraphics[width=3.3in]{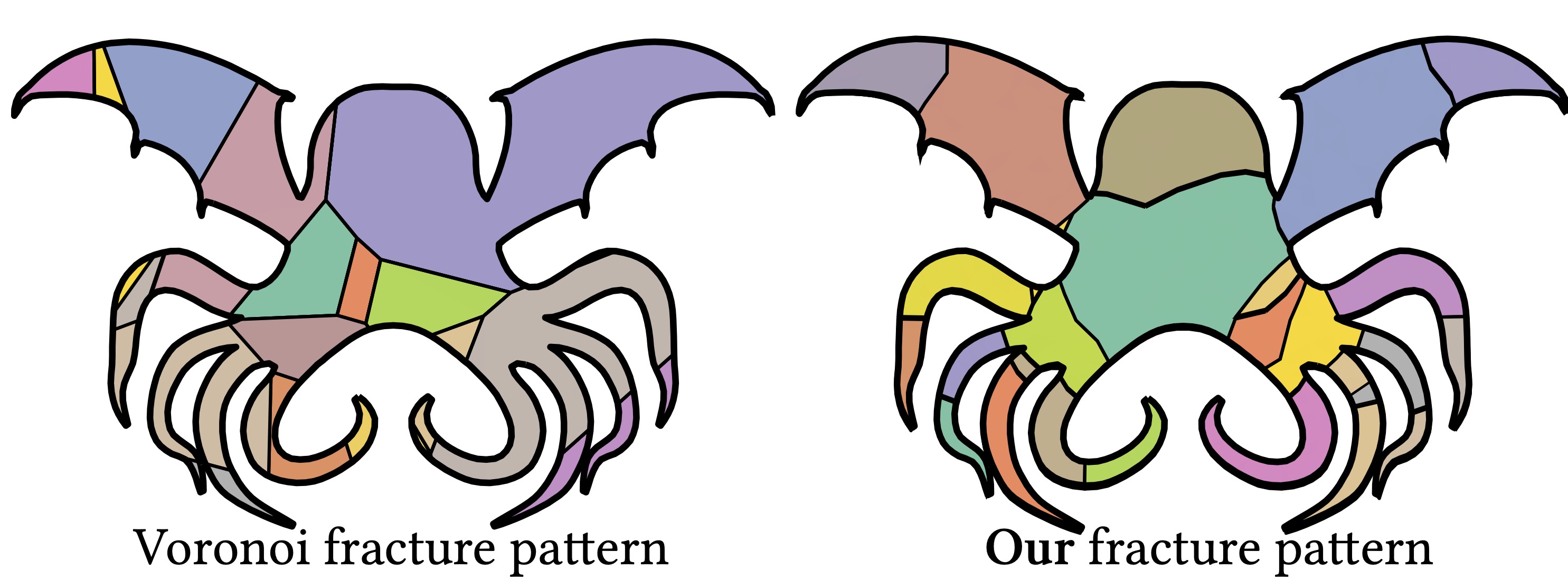}
  \vspace{-0.2cm}
  \caption{\new{Popular Voronoi-based prefracturing results in recognizable,
  unrealistic shapes. Our fracture modes break across weak regions.}
  }
  \label{fig:octopus}
\end{figure}

The patterns and fragmentations formed by an object undergoing brittle fracture
add richness and realism to destructive simulations.
%
%
%
Unfortunately, existing methods for \minor{producing the most high-quality realistic fractures (e.g., for the film industry)}
require hefty simulations 
too expensive for many realtime applications.
An attractive and popular alternative is to rely on precomputed fragmentation
patterns at the modeling stage that can be swapped in at run-time when an impact
is detected.
%
Existing prefracture methods use geometric heuristics that \minor{can} produce
unrealistic patterns oblivious of an object's elastic response profile or
structural weaknesses (see Figs.~\ref{fig:octopus}, \ref{fig:voronoi-bad} and
\ref{fig:sdf-bad}).
Geometric patterns alone also do not answer \emph{which} fragments should break-off
for a given impact at run-time, inviting difficult to tune heuristics or
complete fracture regardless of impact.
As a result,
\minor{these procedural methods find use when fractures are in the background or obscured by particle effects; elsewhere,}
video game studios may rely on artist-authored fragmentation patterns.
%


In this paper, we present a method for prefracturing \minor{stiff brittle materials} which 
draws a direct analogy to a solid shape's elastic vibration modes. 
We compute a shape's \emph{fracture
modes}\footnote{Not to be confused with the ``three modes of fracture'' \cite{irwin1957analysis}.},
which algebraically span the shape's natural ways of breaking apart.
By introducing a continuity objective under a sparsity-inducing norm to the
classic vibration modes optimization problem,
we identify unique and orthogonal modes of fracture in increasing order of a
generalized notion of frequency.

The first $\k$ fracture modes
can \new{be} intersected against each other to define a prefracture pattern as a drop-in
replacement to existing procedural methods
(see \reffig{octopus}).
Furthermore, impacts determined at runtime can be efficiently projected onto the
linear space of precomputed fracture modes to 
obtain impact-dependent fracture without the need for costly
stress computation or crack propagation.

We demonstrate the theoretical and practical \minor{advantages and limitations} of our algorithm over existing
procedural methods and evaluate its accuracy by qualitatively comparing to
existing works in worst-case structural analysis. 
We showcase the benefits of our
algorithm within an off-the-shelf rigid body simulator to produce 
animations on a diverse set of shapes and impacts. We
show the realtime potential of fracture modes with a
prototypical interactive 2D application (see \reffig{interactive} and
accompanying video).

%% file: sections/related.tex
\begin{figure}
\centering
\includegraphics{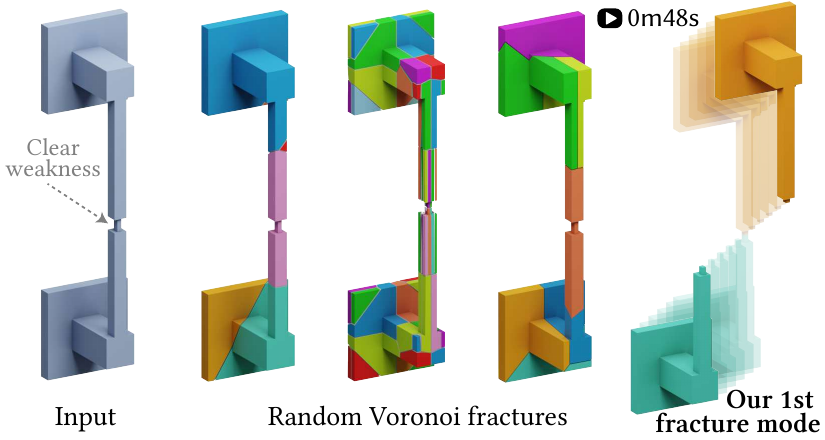}
\vspace{-0.3cm}
\caption{Existing procedural prefracture algorithms (center) rely on randomness
and do not account for the geometrically weak regions of an object, unlike our
proposed fracture modes (right).}
\label{fig:voronoi-bad}
\vspace{-0.5cm}
\end{figure}
\section{Related Work}\label{sec:related}

%
Fracture simulation has an extensive body of previous work;
\citet{muguercia2014fracture} provide a thorough survey.
\minor{Stiff brittle} fracture is characterized when little or no perceptible deformation
occurs before fracture (i.e., when the object is otherwise rigid).
Modeling the dynamic or quasistatic \emph{growth} of brittle fracture patterns
\new{in a high performant way requires not just high spatial resolution but also high temporal resolution at the microsecond scale \cite{kirugulige2007measurement}, modeling stress concentration and subsequent release.}
This process has been approximated, for example, using mass-springs
\cite{norton1991animation,hirota1998generation}, finite elements
\cite{o1999graphical,kaufmann2009,wicke2010dynamic,Pfaff2014ATC,Koschier2015},
boundary elements \cite{Zhu2015,HahnW15,HahnW16}, and the material-point method
\minor{\cite{wolper2019cd,wolper2020anisompm,fanMPM}}.
While any such method can eventually meet realtime demands by lowering the
discretization fidelity (e.g., on low-res cage geometry \new{\cite{parker2009real,muller2004physically}} or
a modal subspace \cite{glondu2012real}) or assuming large enough computational
resources, we instead focus our attention to previous methods which achieve
realtime performance via the well established workflow of prefracturing.
This workflow sidesteps computationally expensive and numerically fragile
remeshing operations. It fits tidily into the existing realtime graphics
pipeline, where geometric resolutions and computational resources can be
preallocated to ensure low latency and consistent performance.

Creating prefracture patterns manually requires skill and time, precluding fully
automatic pipelines.
%
Many commercial packages (e.g., \textsc{Unreal Engine}, \textsc{Houdini})
implement or suggest geometric prefracturing heuristics to segment a shape into
solid subfragments.
Voronoi diagrams of randomly scattered points
\cite{raghavachary2002fracture,oh2012practical} capture the stochastic quality
of fracture, but result in overly regular and convex fragments with perfectly
flat sides. 
\minor{Despite lacking realism, convexity can be advantageous for simulations. For example, it enables realtime collision
detection and even offline at massive scales \cite{zafar2010destroying}. Beyond collision detection, approximate convex decomposition has been employed as
a prefracturing technique, allowing fracture patterns to be applied locally at low cost \cite{muller2013real}}. 
\new{\citet{muller2013real} \minor{rely} on manual intervention at multiple stages, making it not comparable as an automatic method; however, we compare to the Voronoi decomposition it is based on in Figs.~\ref{fig:octopus} and \ref{fig:voronoi-bad}.} 

\begin{figure}
\centering
\includegraphics{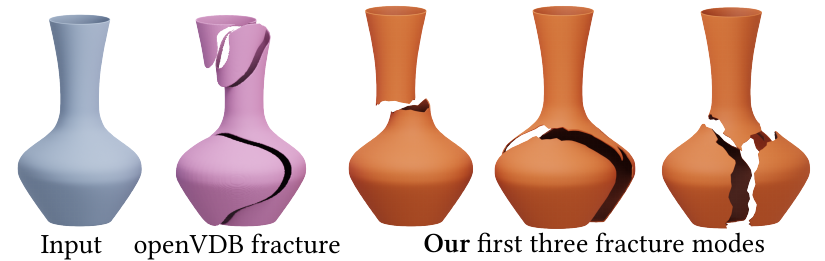}
\vspace{-0.2cm}
  \caption{Level-set methods (e.g., \textsc{OpenVDB} \cite{openvdb}) will
  produce non-convex, yet implausible fracture patterns unrelated to the
  structural integrity of the shape. }
\label{fig:sdf-bad}
\vspace{-0.2cm}
\end{figure}

\new{\citet{SchvartzmanHighDimVoronoi} increase the space of possible
fragments beyond convexity by computing the Voronoi decomposition on a higher
dimensional embedding.} The regularity of these fragments can be augmented
\new{further} by randomly generated ``cutter'' objects (i.e., bumpy planes
slicing though the input shape) or  perturbed level set functions (see, e.g.,
\cite{openvdb}).  
These stochastic methods often miss obvious structural weaknesses (see
\reffig{voronoi-bad}) or result in implausible fragments (see
\reffig{sdf-bad}).
While the defects of these methods can be \minor{resolved manually,} hidden behind destruction dust or
obscured by fast explosions, our fracture modes consistently produce non-convex
fragments whose boundaries originate from minimal stress displacements of the
shape. Beyond taking into account the physical elastic behavior of the geometry,
our method can incorporate constraints to avoid fractures in certain
areas.
\minor{While our fracture modes are slower to compute than geometric-only procedural prefracture, this cost is added only at the offline precomputation stage.}

\new{Given a precomputed fracture pattern, one must decide which fractures are
activated when the object receives a given online impact. Strategies range from
simple heuristics like Euclidean distance thresholds and
centering a spatial fracture pattern on the contact point \cite{muller2013real,su2009energy} to
learning from examples \cite{SchvartzmanHighDimVoronoi}. A key
practical contribution of our fracture modes is that they span a linear subspace
onto which impacts can be cheaply projected to trigger fragment displacements at
runtime, removing the need for heuristics or data-based approaches.}

\begin{figure}
  \centering
  \includegraphics{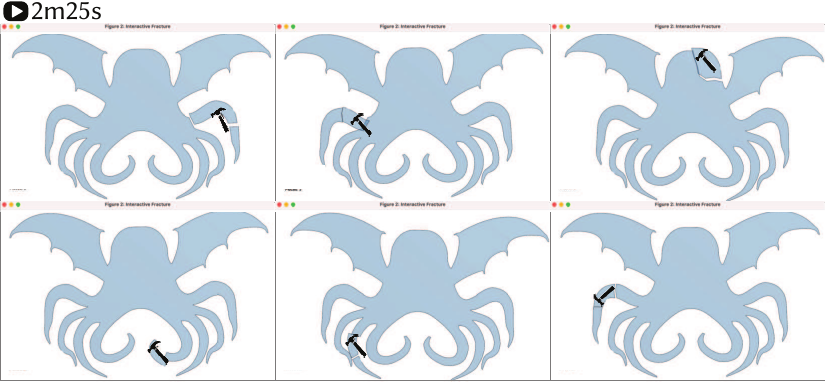}
  \vspace{-0.3cm}
  \caption{Screenshots of our 2D interactive prototype, in which the user can
  select the impact position to obtain different breaking patterns. }
  \label{fig:interactive}
  \vspace{-0.3cm}
\end{figure}

\subsection{Sparsified eigenproblems}\label{sec:eigenproblems} Our approach for
defining fracture modes lies within a broader class of optimization problems of
the form: \new{
\begin{align}
  \argmin_{\X^\top \M \X = \I} \quad \frac{1}{2} \trace{\X^\top \L \X} + \sum_{i=1}^k g(\X_i)\,,\label{equ:genform}
\end{align}}
\noindent where $\X_i$ as the $i^\text{th}$ column of $\X∈ℝ^{n×\k}$ is referred to as a
modal vector or mode, $\M$ and $\L$ are positive semi-definite $n×n$ matrices,
and $g:ℝ^{n}→ℝ$ is a sparsity-inducing norm, like $g(\x) = \|\x\|_{1}$.
\new{When $g \coloneqq 0$, this reduces to the \emph{generalized
eigenvalue problem} (see e.g., \cite{bai2000templates} Chaps. 4-5),} whose
solution satisfies $\L\X =  \M \X Λ$, where the diagonal $\k × \k$ matrix $Λ$
contains the $\k$ smallest eigenvalues ($Λ_{i,i} = \X_iᵀ\L \X_i$).
For non-trivial $g$, we may continue to consider $λ_i = \X_iᵀL\X_i + g(\X_i)$ as
describing the frequency of the $i^\text{th}$ mode.

\citet{Ozolins_2013} proposed the notion of compressed modes using a
sparsity-inducing $\ell_1$-norm to compute localized (sparse) solutions to
Schr{\"o}dinger’s equation.  \citet{CMM2014} extended this idea to compressed
eigenfunctions of the Laplace-Beltrami operator on 3D surfaces, advocating for
an alternating direction method of multipliers (ADMM) optimization method. While
ADMM's standard convergence guarantees \cite{boyd2010} do not apply to
non-convex problem such as \refequ{genform}, \citet{CMM2014} demonstrate
successful local convergence albeit with dependency on the initial guess and
optimization path.
Replacing the $\X^\top \L \X$ term with a data-term, \citet{Neumann2013} use a
similar ADMM approach to create \emph{sparse PCA} bases for mesh deformations.

\citet{BH17} further extend this line of smooth, sparse modal decompositions by
considering $L$ to be the Hessian of an elastic energy. 
They propose an iterative mode-by-mode optimization. The current mode is
optimized by sub-iterations consisting of a quadratic program solve resulting
from linearizing the constraints around the current iterant interleaved with
normalization in order to approach a unit-norm vector.
Despite the conspicuous downside that any sub-optimality of earlier modes is
\emph{locked in} possibly affecting the accuracy of later modes, this method
enjoys performance and robustness improvements over the ADMM approach of
\citet{CMM2014}.
%
Therefore, we follow suit with a similarly mode-by-mode fixed-point iteration
approach.
Unique to our method is that we do not consider the sparsity of the modal vector
itself, but rather the sparsity of the mode's \emph{continuity} over the domain.

%% file: sections/method.tex
\section{Fracture Modes}
  \label{sec:continuousmodes}

%

Given an elastic solid object $\Omega\subset\mathbb{R}^d$ and a deformation map
${u:\Omega\to\mathbb{R}^d}$, we can formulate the object's total strain energy
as
\begin{align}\label{equ:strainenergy}
        E_{\Psi} (u) = \int_{\Omega}\Psi (u,x)dx \minor{,}
\end{align}
where $\Psi$ is the strain energy density function evaluated at points
$x\in\Omega$ in the \textit{undeformed} object.  

  \wf{}{r}{1.00in}{trim={4mm 2mm 0mm
  4mm},width=1.00in}{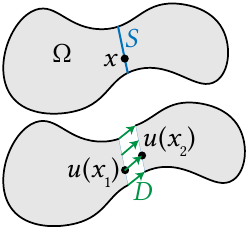}{} 
  \indent
  Suppose we allow the deformation map $u$ to fracture the object $\Omega$ into
  two disjoint pieces $\Omega_1$ and $\Omega_2$ along a \emph{given}
  $(d-1)$-dimensional fracture fault $S$ (see inset). Effectively, we're
  allowing $u$ to be discontinuous at $S$.
  %
  %
  Consider $x_1$ and $x_2$ to be the undeformed positions of points
  infinitesimally on either side of a point $x∈S$ on the fracture fault.
  Then, the difference between $u(x_1)$ and $u(x_2)$ describes the pointwise
  \emph{vector-valued discontinuity} at $x\in S$:
  \begin{align}\label{discontinuitydef}
      D(u,x∈S) = u(x_1) - u(x_2) \in\mathbb{R}^d,
  \end{align}
  where $D = \mathbf{0}$ would indicate continuity or absence of fracture. 
  \new{We can then compute the \emph{discontinuity energy} associated with $S$ as
  \begin{align}\label{equ:discontinuityint}
    \int_{x\in S} \|D(u,x)\|^{2} dx\,,
  \end{align}
  which can be seen as a cohesive surface energy from the FEM crack propagation literature (e.g., \cite{ortiz1999finite}). Unlike similar cohesive energies found in graphics (e.g., for UV mapping \cite{poranne2017autocuts} or shape interpolation \cite{zhu2017planar}), assuming small displacements affords us this simpler, first-order approximation.}

  We now consider that the set of admissible discontinuities is not just a
  single fracture fault, but a finite number of fault patches:
  $S=\{S_1,…,S_p\}$.
  %
  We assume that $S$ comes from a, for now, arbitrary \emph{over-segmentation}
  of $Ω$.
  This could be created with a high-resolution Voronoi diagram, by intersecting
  $Ω$ with random surfaces, voxel boundaries, or by some \emph{a priori}
  distribution of granular subobjects.
  %
  We define the \emph{total discontinuity energy} associated with $u$ as 
  \begin{align}\label{equ:discontinuityenergy}
    E_D(u)=\|D(u,S)\|_{2,1} &\coloneqq \sum_{i=1}^p \sqrt{\int_{S_i} \|D(u,x)\|^{2} dx}.
  \end{align}
  We add this to the strain energy to form the \emph{total energy}:
  \begin{align}\label{equ:totalenergy}
    E(u) = E_{\Psi}(u) + \omega\,E_D(u),
  \end{align}
  where $\omega ∈ ℝ$ is a positive weight balancing the two terms. \new{In the cohesive FEM context, $\omega$ can be understood as the square root of the traction-displacement coefficient in the first fracture phase \cite{chowdhury2000cohesive}}. 
  %
  \new{Minimizing the $\ell_{2,1}$ norm on the matrix $D$ is tantamount to
  minimizing the sparsity-inducing \cite{Candes2008} $\ell_1$ norm on the
  lengths of each row.
  Minimizing the $\ell_1$ norm ($\sum_i \sum_j |D_{ij}|$) directly would also lead
  to sparsity, but the solution would be rotationally dependent.}

\begin{figure}
\centering
\includegraphics{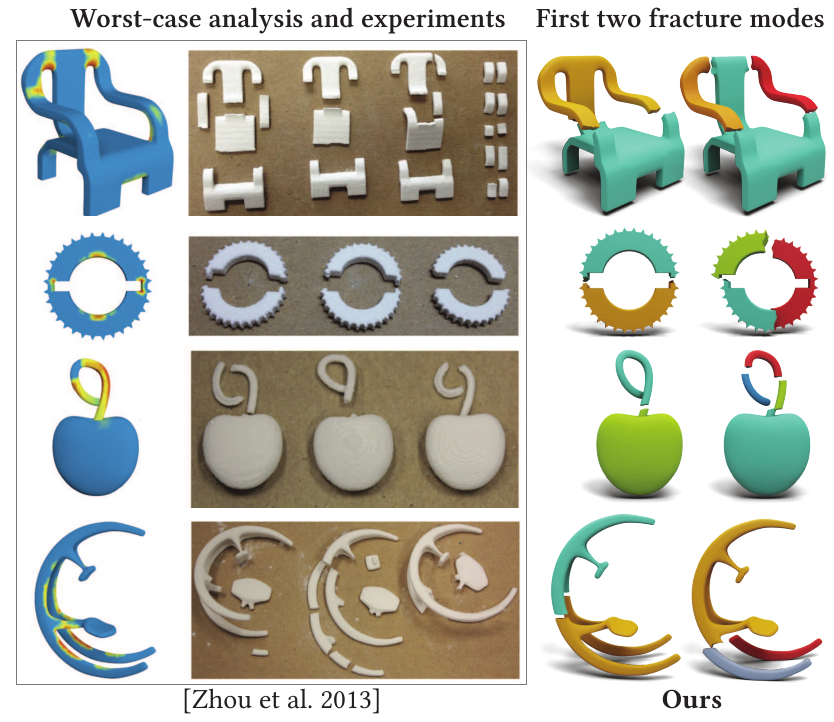}
\vspace{-0.3cm}
\caption{\citet{zhou2013worst} (left) identify the geometrically weakest
  regions of a given shape, which align with their real-world experiments
  (center). Our fracture modes (right) produce fracture patterns qualitatively
  similar.
        }
\label{fig:worst-case}
\vspace{-0.15cm}
\end{figure}

\begin{figure*}
\centering
\includegraphics{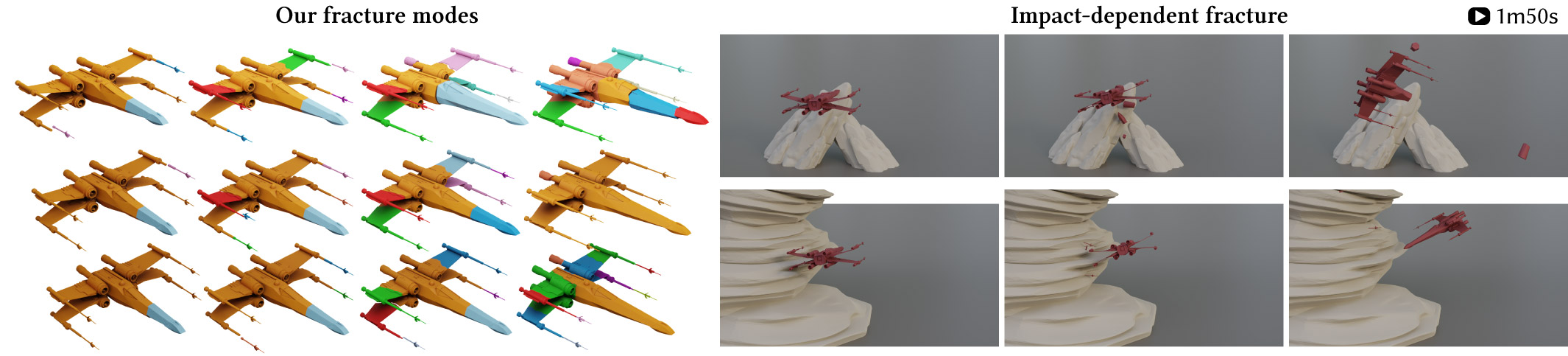}
\vspace{-0.6cm}
\caption{\new{The same prefractured modes can be used to simulate many different
impacts; for example, a racing simulator may destroy the player's spaceship
differently depending on its impact.}}
\label{fig:xwing}
\end{figure*}
  We can now define the $\k$ lowest energy \emph{fracture modes} as the set of
  mass-orthonormal deformation maps $\{u^{i}\}_{i=1}^\k$ that minimize their
  combined total energy; i.e.,
  \begin{equation}\label{equ:fracturemodes}
    \{u^{i}\}_{i=1}^\k = \argmin_{\{u^i\}_{i=1}^\k} \sum\limits_{i=1}^\k E(u^i)\,,\qquad
    \text{s.t.} \quad \int_\Omega (u^i)^{\top} \rho u^j dx = \delta^{i,j},
  \end{equation}
  where $\rho$ is the local mass density and $\delta^{i,j}$ is the Kronecker
  delta.
  In general, for large enough $ω$, minimizers $u^i$ will have exactly zero
  $E_D$ on all but a sparse subset of fault patches in $S$, agreeing with the
  usual sparse coding theory \cite{Candes2008}.

  \subsection{Fracture Modes on Meshes}\label{sec:discretemodes}

  We will derive a discrete formulation of the variational problem in
  \refequ{fracturemodes} for a 2D solid object, represented as a triangle mesh
  $\Omega$ with $n$ vertices and $m$ faces. In this construction, the mesh's $p$
  \emph{interior} edges will correspond to admissible fracture faults
  $S_1,\cdots,S_p$.
  Everything that follows is straightforward to extend to 3D solids by
  exchanging triangles and interior edges for tetrahedra and interior faces.

  \minor{A traditional} piecewise-linear finite element method (FEM) would discretize the
  strain energy $E_{\Psi}$ using hat functions $\varphi_i:\Omega\rightarrow \R,$
  $\forall i=1,\dots,n$ and associate a scalar function $u:\Omega\rightarrow\R$
  with a vector $\u\in\R^n$ such that
  \begin{equation}\label{fem-usual}
    u(x) = \sum_{i=1}^n \u_i \varphi_i(x).
  \end{equation}

  \noindent Vector-valued $u:\Omega\rightarrow \R^d$ such as a deformation would be
  described coordinate-wise in the same way, via a vector $\u\in\R^{dn}$.


  \wf{}{r}{0.87in}{trim={4mm 3mm 0mm 3mm},width=0.87in}{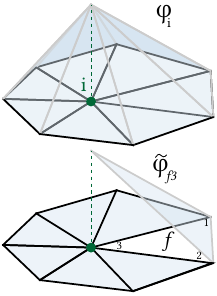}{} 
  \indent
  Hat functions are by construction continuous. Normally, this is a good thing,
  but we would like to have functions with arbitrarily large co-dimension one
  patches of discontinuities.
  %
  Let us introduce the concept of an \emph{exploded} mesh $\tilde{\Omega}$, with
  the same $m$ faces as $\Omega$ and same carrying geometry, but where each
  vertex is effectively repeated for each incident triangle. Thus,
  $\tilde{\Omega}$ is composed of $m$ combinatorially disconnected triangles and
  $3m$ vertices $v_{cf}$ with $c=1,2,3$ and $f=1,\dots,m$.

  Hat functions $\tilde{\varphi}:\tilde{\Omega}\rightarrow \R$ defined on
  $\tilde{\Omega}$ reduce to barycentric coordinate functions extended with zero
  value outside the corresponding triangle (see inset). These trivially span the
  space of piecewise linear scalar \emph{discontinuous} functions $u$ via
  vectors $\u\in\R^{3m}$:
  \begin{equation}
    u(x) = \sum_{f=1}^m \sum_{c=1}^3
    \u_{cf}\tilde{\varphi}_{cf}(x).
  \end{equation}
  We will use this basis for each coordinate of our vector-valued deformation
  map, captured in a vector of coefficients $\u\in\R^{(d+1)md}$, with the
  displacement at vertex $c$ of face $f$ selected by $\u_{cf}\in\mathbb{R}^d$.

  We may now discretize both terms in our energy in \refequ{totalenergy}. First,
  the integral strain energy follows the usual FEM discretization as a sum over
  each element.
  \begin{gather}
      E_{\Psi} (\u)= 
      \sum_{f=1}^m
      \int_{f} \Psi(u,x)dx\,.
  \end{gather}
  \new{We can abstract the choice of $\Psi$ for now by considering small displacements around the rest configuration such that it can be approximated by its Hessian matrix $\Q∈ℝ^{(d+1)md × (d+1)md}$:}
  \begin{gather}
    E_{\Psi} (\u) \approx \frac{1}{2} \u ^\top\Q\u\,.
  \end{gather}
  \new{In \refsec{Q} we discuss a further approximation specific to our model.} 
  Our basis functions $\tilde{\varphi}$ will only allow fractures along the
  mesh's interior edges, therefore, the integral in \refequ{discontinuityenergy}
  breaks into a contribution from each interior edge $e$:
  \begin{equation}\label{equ:integraledge}
    E_e(u) = \sqrt{\int_e \|D(u,x)\|^{2} dx}\,, 
  \end{equation}
  which we can compute exactly by two-point Gaussian quadrature (the integrand
  is a second-order polynomial).

  \wf{}{r}{0.66in}{trim={5mm 3mm 0mm 2mm},width=0.66in}{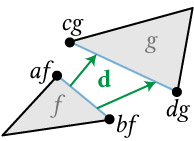}{} 
  \indent
  For a given edge $e$ with length $l$, corresponding to vertex pairs
  $\{af,bf\}$ \emph{and} $\{cg,dg\}$ (see inset), this amounts to
  \begin{align}
    \notag
    E_e(\u) & = \sqrt{\frac{l}{2} \left(\left\|\mathbf{d}\left(\tfrac{+1}{\sqrt{3}}\right)\right\|_2^2 + \left\|\mathbf{d}\left(\tfrac{-1}{\sqrt{3}}\right)\right\|_2^2\right) } 
    \quad
    \quad
    \quad
    \quad
    \quad
    \quad
    \quad
    \quad
    \\
    \intertext{where}
    \mathbf{d}(t) &= 
      \frac{1+t}{2} (\u_{af} - \u_{cg}) + \frac{1-t}{2} (\u_{bf} - \u_{dg})\label{equ:disc}
  \end{align}
  measures the pointwise discontinuity for the quadrature at parametric location
  $t∈[-1,1]$ along the edge.

  The full discontinuity energy associated with the map $u(x)$ is given by
  summing over every interior edge
  \begin{equation} 
    E_D(\u) = \sum_{e=1}^p E_e(\u)\,.
  \end{equation}  


  Finally, we can define the $\k$ lowest-energy \emph{discrete fracture modes}
  as column vectors of a matrix $\U ∈ ℝ^{(d+1)md × \k}$ satisfying
  \begin{equation}
    \label{equ:optim}
          \argmin_{\U^\top \tilde{\M} \U = \I} \quad \frac{1}{2} \trace{\U^\top \Q \U} +
\omega 
          \sum_{i=1}^\k\left( 
    E_D(\U_i)\right) \minor{,}
	\end{equation} 
  where $\tilde{\M}$ is the possibly lumped FEM mass matrix defined on the
  exploded mesh $\tilde{\Omega}$.


%
  %
  \subsection{Optimization}\label{sec:optimization}
  The definition of fracture modes on meshes involves solving the optimization
  problem in \refequ{optim}. While the objective term is convex, the
  orthogonality constraints are not.
  To proceed, we adapt the \emph{Iterated Convexification for Compressed Modes}
  (ICCM) scheme proposed by \citet{BH17}.
  ICCM computes the modes sequentially, assuming the first $i-1$ columns of $\U$
  have been computed. In the original ICCM formulation, the process for finding
  the $i^\text{th}$ column, $\U_i$ proceeds by choosing a random unit-norm
  vector $\c$, then repeatedly solving
  \begin{align}\label{equ:iteration}
      \U_i \leftarrow &
        \argmin_{\u} \frac{1}{2}\u^\top \Q \u + \
          \omega \ E_D(\u)\\
    & \text{subject to} \ 
    \begingroup
\renewcommand*{\arraystretch}{1.25}
    \begin{bmatrix}\U_1^\top \\ \vdots \\ \U_{i-1}^\top \\ \c^\top \end{bmatrix} \tilde{\M} \u = \begin{bmatrix}0\\\vdots\\0\\1\end{bmatrix}
      \endgroup
    \intertext{and updating}
    \minor{\c \leftarrow} & \minor{\frac{\U_i}{\sqrt{\U_i^\top\tilde{\M} \U_i}},}
  \end{align}
    until convergence
  is detected by $\|\U_i - \c\|$ falling below some tolerance $\varepsilon$. 
  By linearizing the (quadratic) norm constraint, the minimization problem
  \refequ{iteration} is a convex conic problem and solved with off-the-shelf
  techniques (see \refapp{conic}).

\begin{figure}
    \includegraphics{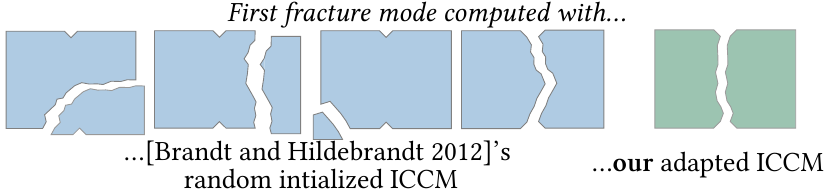}
    \vspace{-0.3cm}  
    \caption{\minor{We adapt the algorithm suggested by \citet{BH17} to use the
      eigenmodes of $\Q$ as initial guesses as opposed to random vectors.}}
      \vspace{-0.2cm}
      \label{fig:iccm}
  \end{figure}

  We found that random initializations for $\c$ not only introduce
  non-determinism, but can also sometimes result in a large number of inner
  iterations and sub-optimal local minima (see \reffig{iccm}).
  %
  Instead, when computing $\U_i$ we initialize $\c$ with the $i^\text{th}$
  continuous eigenvectors of $\Q$ (defined on the unexploded mesh). 
  We compute these $\k$ initial vectors at once using the \textsc{SciPy} wrapper for the sparse eigen solver \textsc{Arpack} \cite{ARPACK}.
  %
  We outline our complete fracture mode computation algorithm in
  \refalg{adap-iccm}.   

  %

  \subsection{Impact-dependent fracture}\label{sec:projection} 
\begin{figure}
    \vspace{-0.1cm}
    \includegraphics{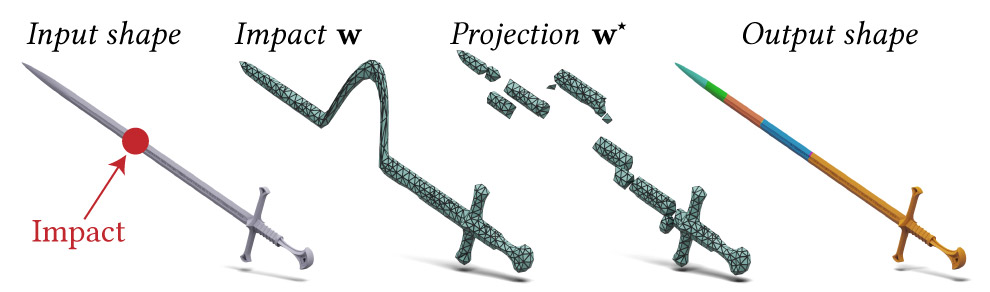}
    \vspace{-0.3cm}
    \caption{\new{We propagate any impact (left) using an elastic shockwave (middle). We then project this propagated impact onto our modes (right). At runtime (see text), we compute $\w^\star$ directly without the need for linear solves.}}
    \vspace{-0.4cm}
    \label{fig:projection-didactic}
  \end{figure}

By construction,
  the columns of $\U$ form an orthonormal basis of the lowest-energy
  $\k$-dimensional subspace of possible fractures of $\Omega$. This key feature
  means we can precompute an object's fracture modes right after its design.
  Then, inside an interactive application we can project any detected impact
  onto our modes to obtain \emph{impact-dependent realtime fracture} (see
  \reffig{projection-didactic}).

  If a collision is detected between $\Omega$ and another object, with contact
  point $p$ and normal $\vec{n}$, we can define the exploded-vertex-wise impact
  vector $\w ∈ ℝ^{(d+1)md}$.
  Ideally, $\w$ would be the displacements determined by an extremely
  short-time-duration simulation of elastic shock propagation. In lieu of being
  able to \new{compute} this in realtime, we use an approximation based on
  distance to smear the impact into the object:
  \begin{equation}\label{equ:impact}
    \w_{cf} = g(p,v_{cf})\,\vec{n}\,,\quad \forall c=1,2,3\,,\,f=1,\dots,m 
  \end{equation}
\new{where $g$ is a filter that vanishes as $v_{cf}$ is far \minor{from} $p$.}
Then, we project $\w$ onto our modes to obtain our
  \emph{projected impact}
  \new{
  \begin{equation}\label{equ:projection}
    \w^{\star} = \sum_{i=1}^\k \U_i \U_i^\top \tilde{\M} \w = \sum_{i=1}^\k \U_i \U_i^\top \tilde{\M} \mathbf{g} \vec{n}
  \end{equation}}

  \begin{figure}
  \includegraphics{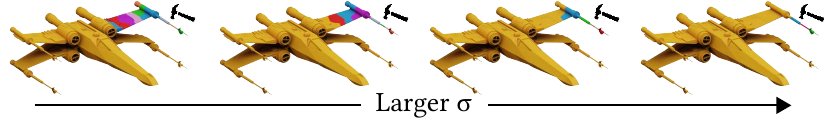}
  \vspace{-0.3cm}
  \caption{\new{A larger threshold (smaller impact) produces less fracture pieces.}}
  \vspace{-0.4cm}
  \label{fig:threshold}
  \end{figure}

\new{An immediate choice of $g$ would be a Gaussian density function centered at $p$. A more physically based choice of $g$ can be obtained through a single implicit timestep of an elastic shockwave equation
\begin{equation}\label{equ:shockwave}
  \mathbf{g} = \left(\tilde{\M}-\tau \tilde{\L}\right)^{-1}\tilde{\M}\delta_p \, ,
\end{equation}
where $\tau$ is the timestep of the simulation our fractures are embedded in. This choice of $g$ has benefits beyond physical inspiration, by ensuring an impact is only blurred onto regions that are geodesically close to one another, regardless of whether they are \emph{Euclideanly} close (see \reffig{projection-didactic}).
However, computing this $\mathbf{g}$ upon impact would involve solving a linear system at runtime. We avoid this by precomputing
\begin{equation}
  \mathbf{A}_i = \U_i^\top\tilde{\M}\left(\tilde{\M}-\tau \tilde{\L}\right)^{-1}\tilde{\M}\,,
\end{equation}}
\new{thus requiring only a matrix multiplication at runtime:
\begin{equation}\label{equ:projection_wave}
  \w^{\star} = \sum_{i=1}^\k \U_i \mathbf{A}_i \delta_p \vec{n}\, .
\end{equation}}

  Let $\tilde{\Omega}_{\w^\star}$ be the exploded mesh $\tilde{\Omega}$ as
  deformed by the map $w^\star$. For any two vertices $v_{af},v_{cg}$ that are
  coincident in $\tilde{\Omega}$ (i.e., that came from the same original vertex
  in $\Omega$), we will \emph{glue} (i.e., \emph{un-explode}) them if their
  deformation maps differ by less than some tolerance, $\|\w^{\star}_{af} -
  \w^{\star}_{cg}\|<\sigma$.  This results in a new \emph{fractured} mesh
  $\Omega^\star$, whose fracture pattern depends meaningfully on the nature of
  the impact and which we can output to the simulation.
  
  \minor{Our single timestep in \refequ{shockwave} is an approximation that makes $\w^\star$ depend linearly on the impact. This has the added effect that scaling $\sigma$ and scaling the magnitude of the impact are equivalent in our model. Thus, $\sigma$ could be linked to the \minor{force} of the impact or the relative speed if one has access to this dynamic information (see \reffig{threshold}). We note that this equivalence is a product of our modeling choices and may not always yield physically accurate results. For example, a large force on a small area may cause immediate local fractures, quickly reducing the stress before it propagates further; in our model, the same impact would likely cause large global fractures.}
  
  \begin{algorithm}[t]
  \DontPrintSemicolon
  Let $\Q$ be a PSD matrix, $\k\in\mathbb{N}$\; $\C \gets
    \text{eigenvectors}(\mathbf{Q},\mathbf{M},\k)$\; \For{$i=1,\dots,\k$}{ $\c
    \gets \C_i$\; \Repeat{$\|\U_i-\c\|\leq\varepsilon$}{ $\U_i\gets$ solve
    \refequ{iteration} (see \refapp{conic})\; $\c \gets \U_i/\sqrt{\U_i^\top
    \tilde{\M} \U_i}$\; } } \Return{$\U$}
    \caption{Fracture Modes via Adapted ICCM}
	\label{alg:adap-iccm}
  \end{algorithm}

  \begin{figure}
    \includegraphics{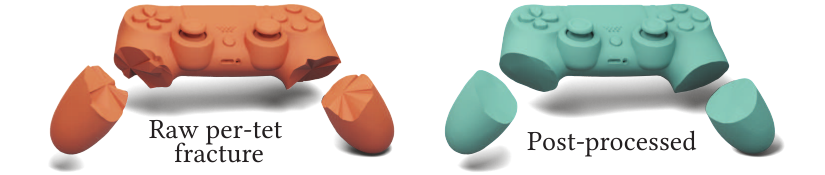}
    \vspace{-0.3cm}
    \caption{We use the upper envelope extraction algorithm by
    \citet{abdrashitov2021interactive} to obtain smoothed fracture faults.}
    \label{fig:rinat}
  \end{figure}

     \subsection{Efficient implementation for real-time
     fracture}\label{sec:3dmethod}
In 3D, our fracture mode computation needs a tetrahedralization of the input's
interior, but practical realtime applications prefer to work with triangle
surface meshes for input and output.
Fortunately, the sparsity inducing discontinuity norm results in fracture modes
which are continuous across most pairs of neighboring tetrahedra.
It is unnecessary to keep the entire tetrahedral mesh at runtime.
Instead, we can determine the connected components determined by neighboring
tetrahedra whose shared face's discontinuity term is below $\sigma$ across all
$\k$ modes \new{(or below the lowest possible $\sigma$ allowed by the dynamic system)}.
The boundary of each component is a \emph{solid} \cite{Zhou16} triangle mesh of
a fracture fragment.
Since the impact projection described above is linear, we can pre-restrict the
projection to vertices on the boundary of these fragments, discarding all
internal vertices and the tetrahedral connectivity.

\subsection{Simple Nested Cages}\label{sec:lazy-cages}
In practice, the input model may be very high-resolution, not yet fully modeled
when fractures are precomputed, or too messy to easily tetrahedralize.
Like many simulation methods before us, we can avoid these potential
performance, workflow and robustness problems by working with a
tetrahedralization coarse cage nesting the input model.
The \textsc{Nested Cages} method of \citet{SachtVJ15} produces tight fitting
cages, but suffers long runtimes, potential failure, and may result in a surface
mesh which causes subsequent tetrahedralization (e.g., using \textsc{TetWild}
\cite{HuZGJZP18} or \textsc{TetGen} \cite{Si15}) to fail.

\wf{}{r}{0.72in}{trim={5mm 3mm 0mm 6mm},width=0.72in}{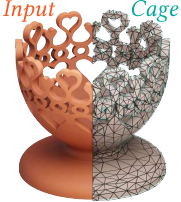}{} 
\indent
Therefore, in \refalg{lazy-cage} we introduce a very simple caging method
inspired by the level-set method of \citet{BenChenWG09}. As an example,
\textsc{Nested Cages} crashes after a minute on the input mesh in the inset,
while our simple algorithm produces a satisfying output cage after 85 seconds.

\begin{algorithm}[h!]
  \DontPrintSemicolon Let $V_\text{in},F_\text{in}$ be the vertex and face lists
  of the input mesh, and\; $m_\text{target}$ the desired number of output
  faces.\;
\SetKwBlock{Block}{}{end} \Block(\textbf{Binary Search} on offset amount $d$) {
  $V_\text{mc}$,$F_\text{mc} \gets$ marching-cubes( distance to
  $V_\text{in}$,$F_\text{in}$ minus $d$)\; $V_\text{d}$,$F_\text{d} \gets$
  decimate $V_\text{mc}$,$F_\text{mc}$ to $m$ faces\; $V_\text{u}$,$F_\text{u}
  \gets$ self-union $V_\text{d}$,$F_\text{d}$ via \cite{Zhou16}\;
  $V_\text{t}$,$T_\text{t},F_\text{t} \gets$ tetrahedralize
  $V_\text{u}$,$F_\text{u}$ via \cite{Si15}\; \If{any step failed
  \textbf{\emph{or}}  $V_\text{t}$,$F_\text{t}$ does not strictly contain
  \new{$V_\text{in}$,$F_\text{in}$}  }{increase $d$\;} \Else{decrease $d$\;} }
  \Return{$V_\text{t}$,$T_\text{t}$}
  \caption{Simple nested cages via binary search}
      \label{alg:lazy-cage}
\end{algorithm}

Like \textsc{Nested Cages}, the output cage will strictly contain the input, but
also by construction we ensure that this cage can be successfully
tetrahedralized (not just in theory). In a sense, this method provides a
different point on the Pareto frontier of tightness-vs-utility. Each step is a
fairly standard geometry processing subroutine with predictable performance, and
one may even consider using it as an initialization strategy for \textsc{Nested
Cages} to improve tightness in the future. We run a max of 10 search iterations,
lasting between 5 and 20 seconds each in our examples.

Fracture modes and solid fragment components on the cage's tetrahedralization
can be transferred to the true input geometry by intersecting each connected
component against the input mesh. In this way, the exterior surface of each
fragment component is exactly a subset of the input mesh.
%

\subsection{Smoothing internal surfaces}\label{sec:rinat}
By our construction, the fracture boundaries will follow faces of the
tetrahedral mesh used for their computation.
This reveals aliasing with frequency proportional to the mesh resolution.
We may optionally alleviate this by treating each extracted per-tet component
membership as a one-hot vector field, which we immediately average onto
(unexploded) mesh vertices stored as a matrix $\Z ∈ [0,1]^{n ×
|\text{components}|}$, so that $\Z_{i,j} ∈ [0,1]$ is viewed as the likelihood
that vertex $i$ belongs to component $j$.
We apply implicit Laplacian smoothing with a time step of $λ$ to columns of
$\Z$:
\begin{equation}\label{equ:smoothing}
  \Z \leftarrow (\M + λ \L)^{-1} (\M \Z),
\end{equation}
where $\M$, $\L$, are the mass and Laplacian matrices, respectively.
The resulting $\Z$ continue to contain fractional values in $[0,1]$
corresponding to a smoothed likelihood.
We now \emph{re}-extract piecewise-linear (triangle mesh) component boundaries
by computing the upper-envelope (tracking the argmax) using the implementation
of \citet{abdrashitov2021interactive}.
While essentially still using the same tetrahedral mesh, utilizing smoothing and
piecewise-linear interpolation greatly reduces aliasing artifacts (see
\reffig{rinat}).

\new{By the nature of the Laplacian, \refequ{smoothing} will push our fracture faults towards smooth surfaces. This is in alignment with our modeling decisions at the beginning of \refsec{continuousmodes}: as our set of possible fault patches $S_i$ becomes larger, the area integral in \refequ{discontinuityenergy} will encourage smoother fracture fault surfaces. Thus, the postprocessing described here is not a departure from our model; rather, a way of alleviating the error introduced by the mesh discretization.}

\new{In the real world, crystalline materials do break along smooth surfaces aligned with their internal structure in a phenomenon known as \emph{cleavage} (see e.g., \cite{ford1932textbook} Part II.I.277). On the other hand, materials like wood or clay do not necessarily break along smooth faults like those produced by our method. This is a well-studied limitation we share with all mesh-based fracture algorithms and which could be alleviated by borrowing strategies from the literature like the Adaptive Fracture Refinement by \citet{ChenAdaptiveFractureRefinement}, \minor{perturbation of crack surface vertices as described by \citet{fanMPM}}, or the use of pre-authored ``splinters'' suggested by \citet{parker2009real}.}

\subsection{Choice of strain energy}\label{sec:Q}
%
%
So far the only requirement on the strain energy density $\Psi$ is that we can
construct its second-order approximation near the rest configuration represented
by the (positive semi-definite) Hessian matrix $\Q$.
We now investigate the effect of choices of $\Q$ and in particular the
relationship with the balancing weight $\omega$.

  \begin{figure}
    \vspace{-0.2cm}
      \includegraphics{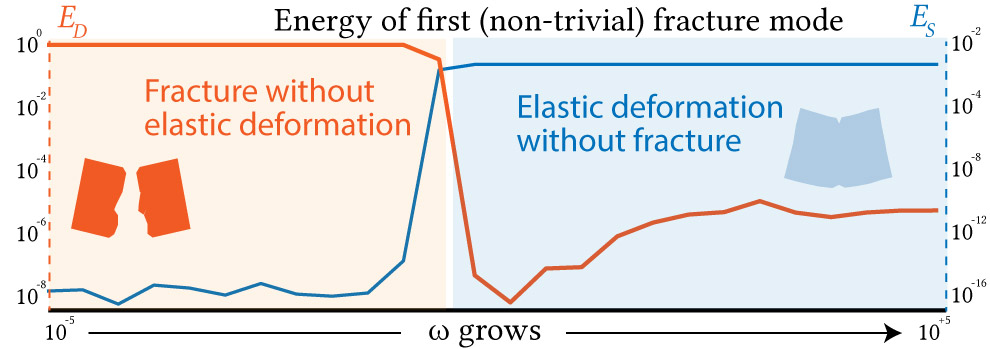}
      \vspace{-0.3cm}
      \caption{As our energy weight changes, our output modes sharply transition
      from only deforming to only fracturing. This gives us the additional
      insight that fractures occur only within the nullspace of the strain
      energy, and has the additional effect of making $\omega$ a simple
      parameter to set.}
      \vspace{-0.5cm}
      \label{fig:energy-plot}
  \end{figure}

To make our investigation concrete, take $\Psi$ to be the linear elastic strain \minor{energy}
density, so that $\Q$ is the common linear elasticity stiffness matrix. \new{Our observations also follow if one chooses $\Q$ to be the Hessian of other, nonlinear energies like the Neohookean or St. Venant-Kirchhoff ones.}
By sweeping across values of $\omega$ we see a sharp change in the first (and
all) fracture mode's behavior with the discontinuity energy dominating over the
strain energy and then sharply swapping (see \reffig{energy-plot}).
When the discontinuity energy is effectively zero, then we have simply recovered
the usual linear elastic vibration modes (albeit in a convoluted way).

\wf{}{r}{1.1in}{trim={4mm 2mm 0mm
  3mm},width=1.1in}{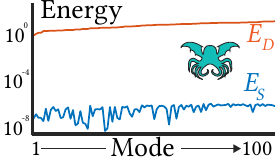}{} 
\new{When the strain energy is effectively zero, then we not only start to see sparse
fractures, but we also see that each fracture fragment undergoes its own
zero-strain energy transformation. This behaviour is consistent even in larger order modes (inset).}
That is, each fragment undergoes a \emph{linearized} rigid transformation, the
only motions in the null space of the strain energy.
\new{Physically, this behaviour naturally aligns our fracture
modes with the traditional definition of} \minor{stiff} \new{\emph{brittle} fracture, where
materials do not significantly deform before breaking.}
\new{This is also interesting from a numerical perspective} as it implies
that the precise choice $\Q$ is irrelevant and only its null space matters.

\begin{table}
    \centering
    \caption{\new{Performance details for all our examples}}\label{table:timing}
    \vspace{-0.3cm}
    \setlength{\tabcolsep}{5.4pt}
\begin{tabularx}{\linewidth}{cccccc}
        \toprule
	Fig. &  \#T & Time/mode (s) & $k$ & $p$ & Impact Proj. (ms) \\
\rowcolor{derekTableBlue}
 \ref{fig:xwing}& 6316 & 2.86 & 30 & 47 & 1.06 \\
\ref{fig:painted-omega}& 3931 & 2.20 & 20 & 117 & 1.21 \\
\rowcolor{derekTableBlue}
\ref{fig:plane} (a)& 3545 & 0.63  & 10 & 35 & 1.13 \\
\ref{fig:plane} (b)& 4993 & 2.63 & 25 & 34 & 1.08 \\
\rowcolor{derekTableBlue}
\ref{fig:skull}& 12162 & 11.6 & 10 & 31 & 1.19 \\
\ref{fig:glass}& 8802 & 5.91 & 15 & 152  & 1.96 \\
\bottomrule
    \end{tabularx}
    \smallskip
\end{table}

With this in mind, we consider whether all linearized rigid transformations
should be admissible.
Since we ultimately care about the fracture pattern created by the modes, we
observed \emph{qualitatively} that the scaling induced by linearized rotations
resulted in small elements breaking off and expanding between fragment
boundaries to reduce the discontinuity energy. 
Rather than attempt to identify these as outliers, we found a simpler solution
is to work with a strain energy that only admits translational motions in its
null space, namely, $\Psi (u,x) = \|\nabla u(x)\|^2$.
%
%
The Hessian of $ \|\nabla u(x)\|^2$ is simply the cotangent Laplacian matrix
$\tilde{\L}∈\R^{(d+1)m×(d+1)m}$ repeated for each spatial coordinate:
  \new{
  \begin{equation}\label{equ:laplacian}
    \Q = \I_d \otimes \tilde{\L}.
  \end{equation}}
This choice of $\Q$ is used in all our examples. 

\paragraph{\new{Efficient precomputation}} 
\new{Our observation regarding the nullspace of $\Q$ can be further exploited to greatly reduce the cost of our offline mode precomputation step. Our strain energy being numerically zero in all our modes means all (exploded) vertices belonging to a single element undergo identical deformations. Therefore, by transforming this observation into an assumption, we may store deformations solely at elements, reducing our number of variables by a factor of $d+1$. This ensures that the strain energy measure on the exploded mesh will always be null, which also means we can remove the quadratic term $\u^\top \Q\u$ from \refequ{iteration}. Further, allowing only per-element deformations also makes our vector discontinuity $D$ necessarily constant along element boundaries, which makes its integral in \refequ{disc} trivial without the need of quadrature nodes. The combination of all these observations significantly reduce the size of our conic problem (see \refapp{conic-smart}), allowing computation of identical fracture modes several orders of magnitude faster.}

%% file: sections/results.tex
\begin{figure}
  \vspace{-0.2cm}
  \centering
  \includegraphics{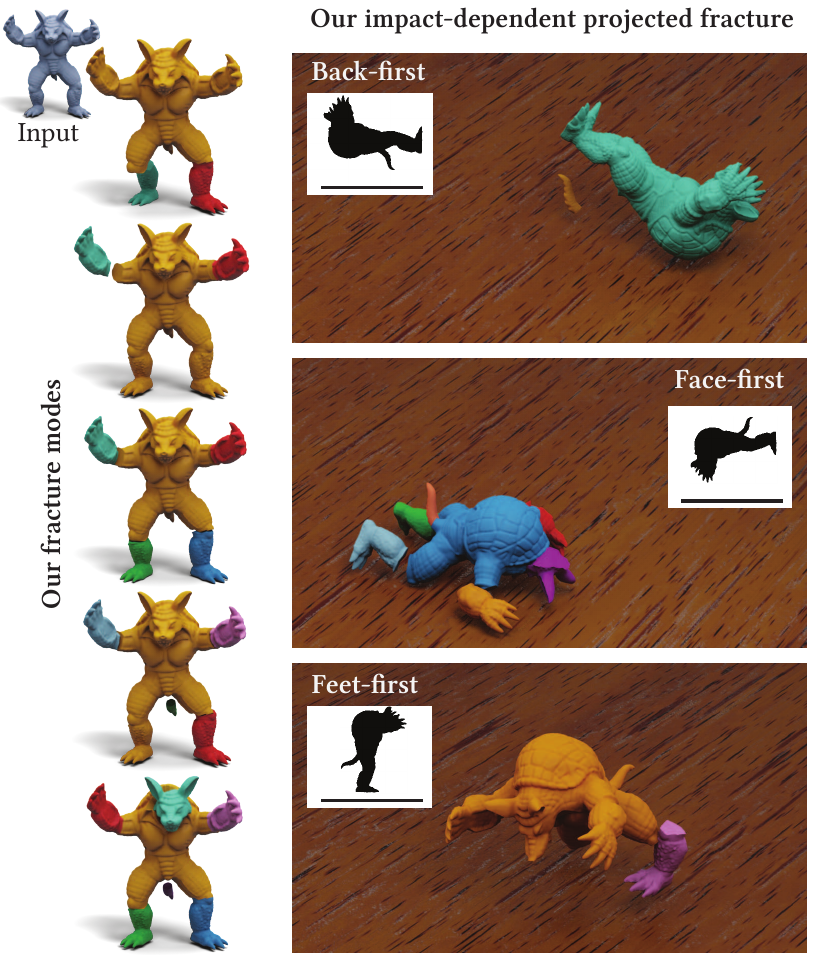}
  \vspace{-0.3cm}
  \caption{Our precomputed fracture modes identify the geometrically weakest
  regions of a shape, and are activated or not on runtime depending on the nature of the
impact.}
  \label{fig:armadillo}
\end{figure}
\section{Timing \& Implementation details}\label{sec:results}

We have implemented our main prototype in \new{Python}, using
\new{\textsc{Libigl} \cite{libigl}}. We used \textsc{Mosek} \cite{mosek} to
solve the conic problem in \refequ{iteration}. We report timings conducted on a
2020 13-inch MacBook Pro with 16 GB memory and 2.3 GHz Quad-Core Intel Core i7
processor. To produce our animations, we follow a traditional \textsc{Houdini} \cite{houdini}
fracture simulation workflow, exchanging the usual Voronoi or openVDB
fracture nodes for our own fractured meshes. Our impact projection step could be
fully integrated into Houdini's rigid body simulator at a minimal performance
cost. Only for simplicity in prototyping, we chose not to do this and instead
compute our final mesh $\Omega^\star$ in Python taking into account the animation's impact
and load it into \textsc{Houdini}
directly to show a prototype of what our algorithm would look like integrated in
a rigid body simulation. 


\new{
Our algorithm's only parameters are the tolerances $\varepsilon$ and
$\sigma$, which we fix at $\varepsilon = 10^{-10}$ and $\sigma =
10^{-3}$. 
As we discuss in \refsec{Q}, a scalar 
$\omega$ will not actually have an effect in the output as long as it is small
enough for us to be within the zero-deformation fracture realm (see
\reffig{energy-plot}).}

Our proposed algorithm works in two steps. First, we precompute a given shape's
fracture modes. This step takes place offline, following \refalg{adap-iccm}. The
computational bottleneck of this section of our algorithm is the conic solve
detailed in \refequ{iteration}. Each mode takes between 0.5 and 12 seconds to compute in our meshes, which have between 3,000 and 15,000 tetrahedra.

Secondly, our impact projection step as detailed in \refsec{projection} is the only
part of our algorithm that happens at runtime. The complexity of this step is
dominated by the projection step in \refequ{projection}, which is
$\mathcal{O}(\k \tilde{n})$, where $\k$ is the number of precomputed modes and $\tilde{n}$ is the
number of vertices in the boundary of the connected components described in
\refsec{3dmethod}. All other elements of our
projection step are $\mathcal{O}(p)$, where $p$ is the number of connected
components (in our example between 10 and 500) and $p<<n$ so they can be disregarded from
the complexity discussion.
In our examples,
$\tilde{n}$ is
between 1,000 and 10,000 and we compute between $\k=20$ and $\k=40$ fracture modes, 
meaning our full runtime step requires between $0.1$ and $1$ million floating point
operations, putting it well within realtime requirements, even if one greatly
increases $\tilde{n}$, $\k$ or the number of objects on scene (note this
projection step only needs to be \new{run} when a collision is detected, and not at
every simulation frame). Our unoptimized, CPU implementation takes
between one and two miliseconds to carry out this step on our laptop (see Table \ref{table:timing}). 

\begin{figure}
\centering
\vspace{-0.3cm}
\includegraphics{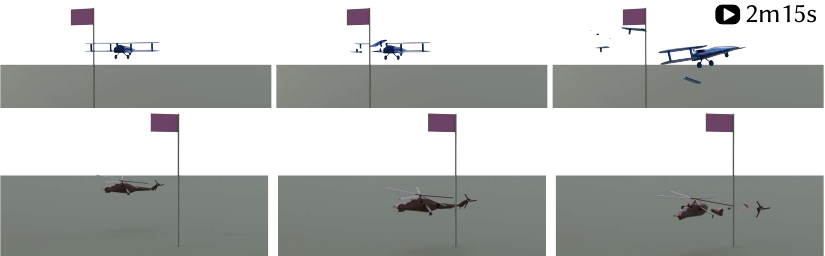}
\vspace{-0.3cm}  
\caption{\new{Similar impacts result in different fracture patterns once we have
  computed our fracture modes for different objects.}} 
\label{fig:plane}
\end{figure}

\begin{figure}
  \vspace{-0.4cm}
  \centering
  \includegraphics{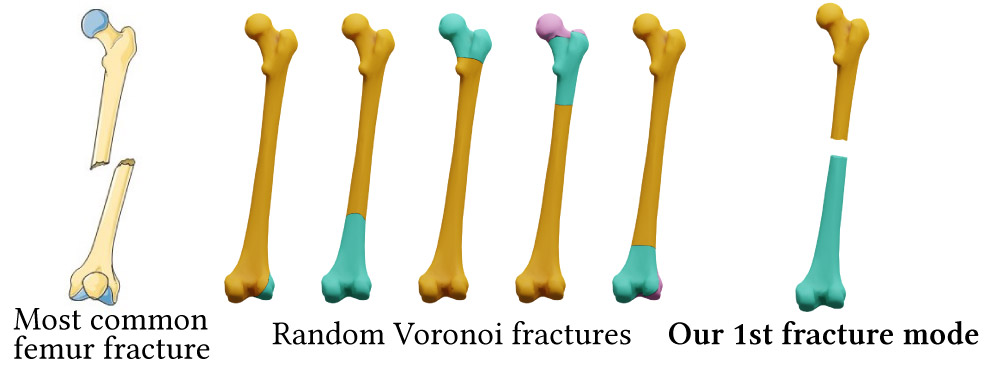}
  \vspace{-0.3cm}
  \caption{A healthy femur will usually break at the mid-shaft \new{(see e.g., \cite{adnan2012frequency})}, as our first
  fracture mode correctly identifies, unlike Voronoi-based algorithms. \new{Left image by Servier Medical Art under CC BY-SA 3.0.}}
  \label{fig:bone}
  \vspace{-0.3cm}
\end{figure}

\begin{figure*}
  \vspace{-0.2cm}
\centering
\includegraphics{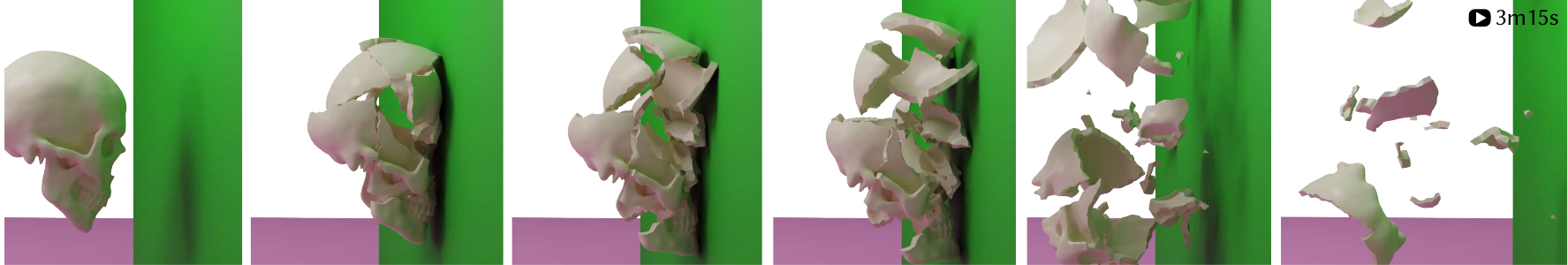}
\vspace{-0.6cm}
\caption{\new{Ouch, my head hurts!}}
\vspace{-0.2cm}
\label{fig:skull}
\end{figure*}

\begin{figure}
  \centering
  \vspace{-0.4cm}
  \includegraphics{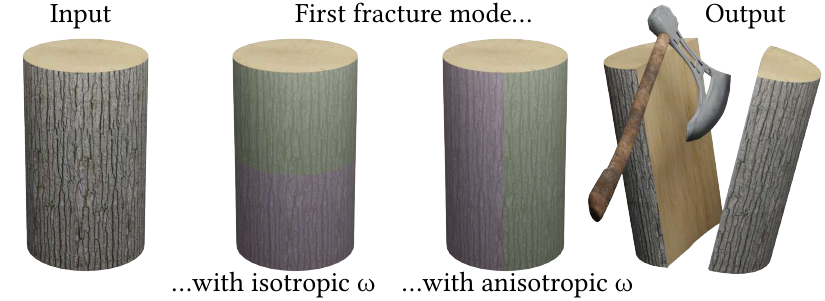}
  \vspace{-0.4cm}
  \caption{\minor{We can simulate the natural breaking tendencies of anisotropic materials like wood through weights in our vector valued discontinuity.}}
  \label{fig:wood}
\end{figure}

\section{Experiments \& Comparisons}\label{sec:experiments}

Our proposed fracture modes naturally identify the regions of a shape that are
geometrically weak, as opposed to existing procedural prefracture algorithms. We
make this explicitly clear in Figs. \ref{fig:voronoi-bad} and \ref{fig:bone}, where existing prefracture
work fails to identify even the most obvious intuitive breaking patterns which
are present in our first (non-trivial) fracture mode. Even in less didactic
examples, Voronoi-based prefracture methods result in convex, unrealistic and easily
recognizable pieces (see Figs. \ref{fig:prefracture} and \ref{fig:octopus}), while our
fracture modes are realistic and can produce a much wider set of shapes.

``Realism'' in a fracture simulation is a hard quantity to evaluate; however,
there exist works on structural analysis like \cite{zhou2013worst} that idenfity the weakest regions of a
given object. In \reffig{worst-case}, we show how our fracture modes produce
breaking patterns that align both with their analysis as well as with their
real-life experiments.

\minor{We model heterogeneous and anisotropic materials by incorporating a vector field $\eta:\Omega\rightarrow\mathbb{R}^d$ to the discontinuity energy:
\begin{equation}\label{equ:discontinuityenergyanisotropic}
    E_D(u)= \sum_{i=1}^p \sqrt{\int_{S_i} \| \eta(x) \circ D(u,x)\|^{2} dx}\,,
 \end{equation}
where $\circ$ denotes Hadamard (elementwise) multiplication.
In \reffig{painted-omega}, we experiment with varying the magnitude of $\eta$ as an artist control tool to designate regions that should not fracture. In \reffig{wood}, we make $\eta = (10,10,1)$ to favour vertical faults over horizontal ones.
}




\subsection{Fracture simulations}\label{sec:fracture-simulation}

Our proposed method is ideal for its use in interactive applications. In
\reffig{interactive}, we show screenshots of our 2D realtime fracture
interactive app. The user can cause different impacts on the object and see the
fracture patterns that result from them by projecting onto our fracture modes.

The interactive Computer Graphics application \emph{par excellence} is video
games. In
\reffig{xwing}, we show a prototype where our precomputed fracture modes for a
Space Wizard Vehicle can be
stored so that the player sees different fracture behaviours depending on the
received impact. In \reffig{plane}, we precompute the fracture modes for two different
vehicles and show how they break under a similar impact.

\begin{figure}
  \vspace{-0.2cm}
  \centering
  \includegraphics{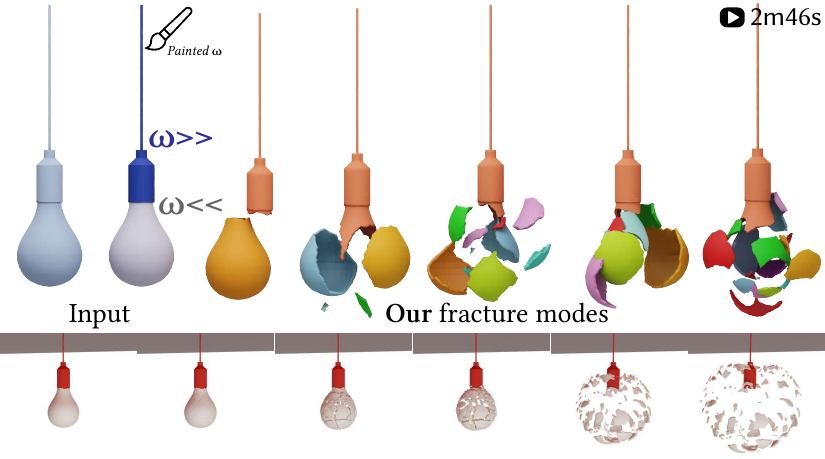}
  \vspace{-0.4cm}
  \caption{The magnitude of our \minor{$\eta$} parameter can be painted into our input to signal areas
  that shouldn't fracture; e.g., in objects with different materials.
  }
  \label{fig:painted-omega}
\end{figure}

\begin{figure*}
\centering
\includegraphics{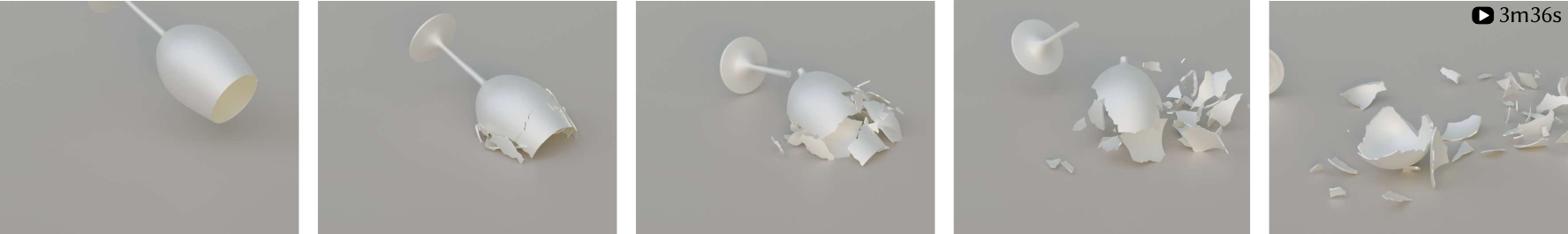}
\vspace{-0.6cm}
\caption{\new{A glass cup shatters, resulting in many non-convex pieces that would be
impossible to obtain with Voronoi-based prefracture methods.}}
\label{fig:glass}
\end{figure*}

Our algorithm can be used for any realtime fracture application, from simple
objects breaking into solid pieces in the foreground of an animation (see
\reffig{armadillo}) to thin shells shattering upon impact (see Figs. \ref{fig:glass} and \ref{fig:painted-omega}). In \reffig{skull}, 
\new{we use our fracture modes to simulate a human skull breaking into many pieces upon impact with a wall.}

\begin{figure}
  \vspace{-0.2cm}
  \includegraphics{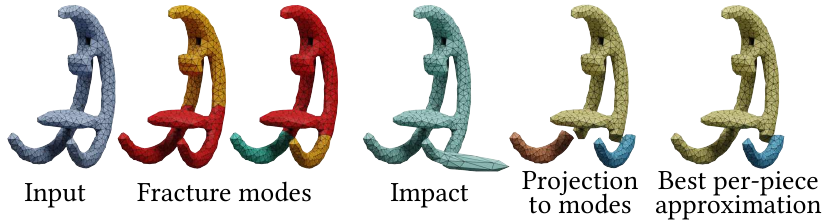}
    \vspace{-0.3cm}
    \caption{\new{Our modes' global nature (left) means some regions can be artificially linked (right middle). We could solve this by exchanging our mode projection for the least-squares best constant-per-piece approximation (right)}}
    \label{fig:lsqr}
    \vspace{-0.3cm}
\end{figure}


%% file: sections/conclusion.tex
\section{Limitations \& Future Work}\label{sec:conclusion}
%

Our fracture modes method is intended for \minor{stiff} brittle fracture. We conjecture that \minor{general rigid fracture and even}
ductile fracture simulation could also benefit from our sparse-norm formulation.
In future work, we would like to improve the performance of our precomputation
optimization. We experimented with \textsc{Manopt} \cite{manopt}, but so far
observed significantly slower performance than our proposed method.
%
For very large meshes, the projection step could exceed CPU usage allowances for
realtime applications. It may be possible to conduct this entirely on the GPU.
%

\new{Our fracture modes are global in nature, meaning they create relations between regions of the object that will not typically fracture together \minor{(unlike other prefracture methods like \cite{oh2012practical})}. A way of preventing a fracture in one location from also causing an undesired fracture elsewhere is to use our modes only to identify the pieces that could break off of an object in the precomputation step, and swap our realtime impact projection for a least squares constant-per-piece approximation (see \reffig{lsqr}).}

\new{
  Our use of an \emph{exploded mesh} $\tilde{\Omega}$ allows us to expand the usual finite element hat function basis to include discontinuities along element boundaries. 
  \minor{This mesh dependency is not present in traditional Voronoi or plane-cutting prefracture algorithms, and can lead to visible artifacts if the simulation mesh is too coarse.}
  We alleviate it with \emph{post-facto} smoothing (\refsec{rinat}). Another way of reducing it (at a performance cost) would have been to include basis functions with sub-mesh-resolution discontinuities in the style of XFEM \cite{kaufmann2009,chitalu2020displacement}.
}

\new{Our algorithm is designed to fit into realtime rigid body simulations like those encountered in video games. Thus, our outputs will not contain partial fractures (unlike e.g., \cite{muller2013real}).}

\wf{}{r}{0.7in}{trim={4mm 4mm 0mm
  4mm},width=0.7in}{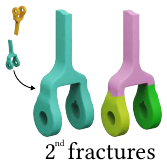}{} 
  \minor{\indent Secondary fractures were not included in our simulations. Computing a new set of fracture modes for each piece would exceed realtime constraints. While one could obtain plausible secondary fractures by  
  restricting our precomputed fracture modes and pattern to each primary fracture piece (see inset), there is no guarantee that these would match the individual piece's fracture modes.}

\new{Our fracture mode computation considers only the magnitude of the vector-valued discontinuity and disregards its direction. Promising future work could include treating tangential and normal components differently as a way of simulating different material properties.}

\begin{figure}
  \vspace{-0.2cm}  
  \includegraphics{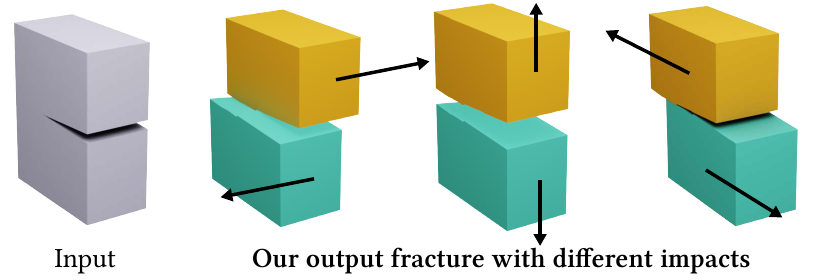}
    \vspace{-0.3cm}
    \caption{\new{If a given fracture mode is contained in our fracture modes, it can occur under any directional impact, as shown in this simple example inspired by \citet{o1999graphical}.}}
    \label{fig:modesoffracture}
    \vspace{-0.3cm}
\end{figure}

\new{Our method belongs to the class of prefracture, not dynamic algorithms. Nonetheless, our method can be evaluated on dynamic fracture benchmarks like the notched block in \cite{o1999graphical}: if a given fracture plane is contained in one of our fracture modes, 
\minor{it can be present in the fractured output (see \reffig{modesoffracture}).
The fracture fault will be the same regardless of the directionality of the impact. This deviates from the real-world mechanical behaviour, where faults will be different for brittle materials under uniaxial tension, pure shear, and torsion loads (see \cite{lawn1993fracture}, Chap. 2).}}

We hope our introduction of fracture eigenmodes inspires the realtime simulation
community further to use the well-studied tools of modal analysis to this rich
problem, and the broader Computer Graphics research community to look at other
open problems with this modal lens.

\begin{figure}
  \vspace{-0.3cm}  
  \includegraphics{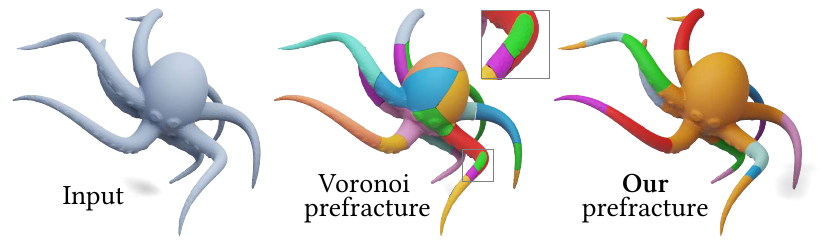}
    \vspace{-0.3cm}
    \caption{The pieces generated by Voronoi-based methods can be extremely
      unrealistic (center). By combining all the possible fractures in all our
      modes into a single \emph{prefractured} mesh, we can provide a
      zero-extra-realtime-cost alternative to procedural algorithms.
  }
    \label{fig:prefracture}
    \vspace{-0.3cm}
\end{figure}

%% file: sections/acks.tex
\begin{acks}
This project is funded in part by NSERC Discovery (RGPIN2017–05235,
RGPAS–2017–507938), New Frontiers of Research Fund (NFRFE–201), the Ontario
Early Research Award program, the Canada Research Chairs Program, the Fields
Centre for Quantitative Analysis and Modelling and gifts by Adobe Systems.
The first author is supported by an NSERC Vanier Canada Scholarship, the FA\&S
Dean's Excellence Scholarship, the Beatrice ``Trixie'' Worsley Graduate
Scholarship and a Connaught International Scholarship. The second, third, fourth and fifth authors were supported by the 2020 Fields Undergraduate Summer
Research Program.

We acknowledge the authors of the 3D models used throughout this paper and thank
them for making them available for academic use: MakerBot
(\reffig{modes-vs-modes}, CC BY 4.0), HQ3DMOD (Figs. \ref{fig:sdf-bad} and
  \ref{fig:glass}, TurboSquid 3D
Standard Model License), Freme Minskib (\reffig{xwing}, CC BY-NC 4.0), 3Demon
(\reffig{projection-didactic}, CC BY-NC-SA 4.0), Reality\_3D (\reffig{rinat}, CC
BY 4.0), Alex (\reffig{plane}, CC BY-NC-SA 4.0), Falha Tecnologica
(\reffig{painted-omega}, TurboSquid 3D Standard Model License), LeFabShop
(\reffig{skull}, CC BY-NC 4.0), The Database Center for Life Science
(\reffig{bone}, CC BY-SA 2.1) and Gijs (inset in \refsec{lazy-cages}, CC BY-NC 4.0).

We would like to thank
Chris Wojtan, David Hahn and Klint Qinami for early experiments and
discussions of sparse-norm fracture models; Eitan Grinspun, David I.W. Levin,
Oded Stein and Jackson Phillips for insightful conversations; Rinat Abdrashitov for providing an
implementation of his algorithm mentioned in
\refsec{rinat}; Qingnan Zhou for providing the 3D models used in
\reffig{worst-case}; Xuan Dam, John
Hancock and all the University of Toronto Department of Computer Science
research, administrative and maintenance staff
that literally kept our lab running during a very hard year.  
\end{acks}

%% file: sections/appendix.tex
\appendix
\section{Canonical conic program form of \refequ{iteration}}\label{app:conic} 

\new{Let us define a sparse matrix $\D \in \mathbb{R}^{pd × d(d+1)m}$ that
operates on a deformation map and evaluates the vector-valued discontinuities at
all the relevant integration quadrature points. The ordering of the rows of $\D$
is arbitrary, and we choose it such that $\D$ can be separated into $d$ blocks, one for each quadrature point. 
}

\new{
For example, in the case $d=2$, we choose said ordering such that the first $p$ rows of $\D
\u$ are the vector-valued edge-wise discontinuity 
\[\sqrt{\frac{l_e}{2}} \mathbf{d}  \left(\frac{1}{\sqrt{3}}\right)\]
from \refequ{disc} and the last $p$ rows of $\D \u$ correspond to
\[ \sqrt{\frac{l_e}{2}} \mathbf{d}  \left(\frac{-1}{\sqrt{3}}\right)\,. \]
}

Since $\Q$ is positive semi-definite, we can write it as $\Q =
\mathbf{R}^\top \mathbf{R}$ for some matrix $\mathbf{R}$. Define
\new{
\begin{equation}
  \Y_e = (\D \u)_e\,,\forall
  e=1,\dots,2p \minor{,}
\end{equation}
}
where $l_e$ is the length of edge $e$. Next, define
\begin{equation}
  \qquad \mathbf{r}_i = (\mathbf{R} \u)_i\,,\forall i=1,\dots,d(d+1)m.\quad
\end{equation}
Then, \refequ{iteration} can be written in the canonical form
\new{
\begin{gather}
     \argmin_{\u,t,\Y,\z}  \begin{bmatrix}
		1 \\
		\zeros \\
		\zeros \\
		\zeros \\
		\displaystyle\ones \\
	\end{bmatrix}^\top
	\begin{bmatrix}
		t \\
		\mathbf{r} \\
		\u \\
		\Y \\
		\z \\
	\end{bmatrix}
     \intertext{subject to}
     t \geq \sqrt{\mathbf{r}_1^2+\hdots+\mathbf{r}_{d(d+1)m}^2}\notag\\
          \z_e \geq \sqrt{ \sum_{s=0}^{d-1}\left\|\Y_{e+sp}\right\|_2^2 } \; \; \forall e=1,\dots,p
          \notag\\
     \Y=\D \u\notag\\
     \mathbf{r}=\mathbf{R} \u\notag\\
          \c^{\top} \tilde{M} \u = 1\notag \\
          \U^{j}\tilde{M} \u= 0\,,\quad\forall j=1,\dots,i-1 \minor{.}\notag 
\end{gather}
}

\vspace{-0.3cm}
\section{Efficient conic program from \refsec{Q}}\label{app:conic-smart}

\new{Let $\C$ be the $\R^{(d+1)m\times m}$ matrix of ones and zeros that transfers values from elements to vertices in the exploded mesh, and let us assume now that we are storing per-element deformations in vectors $\v,\U_j,\c\in\R^{dm}$. Then, our conic problem from \refapp{conic} becomes
\begin{gather}
	\argmin_{\v,t,\Y,\z}  \begin{bmatrix}
	   1 \\
	   \zeros \\
	   \zeros \\
	   \zeros \\
	   \displaystyle\ones \\
   \end{bmatrix}^\top
   \begin{bmatrix}
	   t \\
	   \mathbf{r} \\
	   \v \\
	   \Y \\
	   \z \\
   \end{bmatrix}
	\intertext{subject to}
	t \geq \sqrt{\mathbf{r}_1^2+\hdots+\mathbf{r}_{d(d+1)m}^2}\notag\\
		 \z_e \geq \sqrt{ \sum_{s=0}^{d-1}\left\|\Y_{e+sp}\right\|_2^2 } \; \; \forall e=1,\dots,p
		 \notag\\
	\Y=\D \C \v\notag\\
	\mathbf{r}=\mathbf{R}\C \v\notag\\
	\c^{\top} \tilde{M} \u = 1\notag \\
	\U^{j}\tilde{M} \u= 0\,,\quad\forall j=1,\dots,i-1 \minor{.}\notag
\end{gather}}

\new{By construction, $\Q\C\v=0\Rightarrow \mathbf{R}\C\v = 0$, which means we can remove $t$ and $\mathbf{r}$ as variables entirely. Further, since the vector-valued discontinuity is constant across element boundaries, $\Y_e = \Y_{e+sp}$, with $s=0,\dots,d-1$, meaning that we can remove the summation from the $z_e$ cone and consider only the first $p$ rows of $\Y$. This leads to the equivalent, simpler conic program
\begin{gather}
	\argmin_{\v,\Y}  \begin{bmatrix}
	   \zeros \\
	   \zeros \\
	   \displaystyle\ones \\
   \end{bmatrix}^\top
   \begin{bmatrix}
	   \v \\
	   \Y \\
	   \z \\
   \end{bmatrix}
	\intertext{subject to}
		 \z_e \geq \sqrt{ d\left\|\Y_{e}\right\|_2^2 } \; \; \forall e=1,\dots,p
		 \notag\\
	(\Y)_e=(\D \C \v)_e\;\;\forall e=1,\dots,p\notag\\
		 \c^{\top}\C^\top \M \C\v = 1\notag \\
		 \U^{j}\C^\top \M \C\v= 0\,,\quad\forall j=1,\dots,i-1 \minor{.}\notag
\end{gather}}